\input jytex.tex   
\typesize=10pt \magnification=1200
\baselineskip17truept 
\hsize=6truein\vsize=8.5truein 
\vglue 1truein
\sectionnumstyle{blank}
\chapternumstyle{blank}
\chapternum=1
\sectionnum=1
\pagenum=0

\def\begintitle{\pagenumstyle{blank}\parindent=0pt\begin{narrow}[0.4in]}
\def\endtitle{\end{narrow}\newpage\pagenumstyle{arabic}}


\def\beginexercise{\vskip 20truept\parindent=0pt\begin{narrow}[10 
truept]}
\def\endexercise{\vskip 10truept\end{narrow}}


\def\eql#1{\eqno\eqnlabel{#1}}
\def\ref{\reference}
\def\peq{\puteqn}
\def\pref{\putref}

\def\mgn{\marginnote}
\def\bex{\begin{exercise}}
\def\eex{\end{exercise}}


\def\mbox#1{{\leavevmode\hbox{#1}}}

\def\hspace#1{{\phantom{\mbox#1}}}

\def\al{\alpha}
\def\bPhi{{\bmit\Phi}} 
\def\be{\beta}

\def\La{\Lambda}

\def\det{{\rm det\,}}
\def\Det{{\rm Det\,}}

\def\zf{$\zeta$--function}


\def\frac#1/#2{\leavevmode\kern.1em
\raise.5ex\hbox{\the\scriptfont0 #1}\kern-.1em/\kern-.15em
\lower.25ex\hbox{\the\scriptfont0 #2}}
\def\sfrac#1/#2{\leavevmode\kern.1em
\raise.5ex\hbox{\the\scriptscriptfont0 #1}\kern-.1em/\kern-.15em
\lower.25ex\hbox{\the\scriptscriptfont0 #2}}

\def\gtorder{\mathrel{\raise.3ex\hbox{$>$}\mkern-14mu
             \lower0.6ex\hbox{$\sim$}}}
\def\ltorder{\mathrel{\raise.3ex\hbox{$<$}\mkern-14mu
             \lower0.6ex\hbox{$\sim$}}}

\def\semidirprod{\rlap{\ss C}\raise1pt\hbox{$\mkern.75mu\times$}}
\def\for{\lower6pt\hbox{$\Big|$}}
\def\fish{\kern-.25em{\phantom{abcde}\over \phantom{abcde}}\kern-.25em}


\def\boxit#1{\vbox{\hrule\hbox{\vrule\kern3pt
        \vbox{\kern3pt#1\kern3pt}\kern3pt\vrule}\hrule}}
\def\dalemb#1#2{{\vbox{\hrule height .#2pt
        \hbox{\vrule width.#2pt height#1pt \kern#1pt
                \vrule width.#2pt}
        \hrule height.#2pt}}}

\def\frac#1#2{{{#1}\over{#2}}}


\def\ie{{\it i.e. }}


  %

\def\3j#1#2#3#4#5#6{\left\lgroup\matrix{#1&#2&#3\cr#4&#5&#6\cr}
\right\rgroup}

\def\m?{\mgn{?}}

\def\beq{\begin{eqnarray}}
\def\eeq{\end{eqnarray}}


\def\aop#1#2#3{{\it Ann. Phys.} {\bf {#1}} (19{#2}) #3}

\def\cmp#1#2#3{{\it Comm. Math. Phys.} {\bf {#1}} (19{#2}) #3}
\def\cqg#1#2#3{{\it Class. Quant. Grav.} {\bf {#1}} (19{#2}) #3}

\def\jgp#1#2#3{{\it J. Geom. and Phys.} {\bf {#1}} (19{#2}) #3}
\def\jmp#1#2#3{{\it J. Math. Phys.} {\bf {#1}} (19{#2}) #3}
\def\jpa#1#2#3{{\it J. Phys.} {\bf A{#1}} (19{#2}) #3}

\def\np#1#2#3{{\it Nucl. Phys.} {\bf B{#1}} (19{#2}) #3}
\def\pl#1#2#3{{\it Phys. Lett.} {\bf {#1}} (19{#2}) #3}

\def\prD#1#2#3{{\it Phys. Rev.} {\bf D{#1}} (19{#2}) #3}

\def\cras#1#2#3{{\it Comptes Rend. Acad. Sci. (Paris)} {\bf{#1}} (#2) #3}

\def\mpcps#1#2#3{{\it Math. Proc. Camb. Phil. Soc.} {\bf{#1}} (19{#2}) #3}

\def\am#1#2#3{{\it Acta Mathematica} {\bf {#1}} (19{#2}) #3}
\def\aim#1#2#3{{\it Adv. in Math.} {\bf {#1}} (19{#2}) #3}
\def\ajm#1#2#3{{\it Am. J. Math.} {\bf {#1}} ({#2}) #3}

\def\aom#1#2#3{{\it Ann. of Math.} {\bf {#1}} (19{#2}) #3}

\def\cpde#1#2#3{{\it Comm. Partial Diff. Equns.} {\bf {#1}} (19{#2}) #3}

\def\invm#1#2#3{{\it Invent. Math.} {\bf {#1}} (19{#2}) #3}
\def\ijpam#1#2#3{{\it Ind. J. Pure and Appl. Math.} {\bf {#1}} (19{#2}) #3}
\def\jdg#1#2#3{{\it J. Diff. Geom.} {\bf {#1}} (19{#2}) #3}

\def\jmpa#1#2#3{{\it J. Math. Pures. Appl.} {\bf {#1}} ({#2}) #3}

\def\ojm#1#2#3{{\it Osaka J.Math.} {\bf {#1}} ({#2}) #3}

\def\pja#1#2#3{{\it Proc. Jap. Acad.} {\bf {A#1}} (19{#2}) #3}

\def\tams#1#2#3{{\it Trans. Am. Math. Soc.} {\bf {#1}} (19{#2}) #3}

\begin{title}  
\vglue 1truein
\righttext {MUTP/98/4}
\vskip15truept
\centertext {\Bigfonts \bf On the relevance of the multiplicative anomaly}
\vskip 20truept 
\centertext{J.S.Dowker\footnote{dowker@a13.ph.man.ac.uk}}
\vskip 7truept
\centertext{\it Department of Theoretical Physics,}
\centertext{The University of Manchester, Manchester, England}
\vskip 20truept
\centertext {Abstract}
\vskip10truept
\begin{narrow}
We shed doubt on a commonly used manipulation in computing the partition
function for a matrix valued operator together with the attendant invocation 
of the multiplicative anomaly.
\end{narrow}
\vskip 5truept
\righttext {March 1998}
\vskip 60truept
\vfil
\end{title}

\pagenum=0

In some recent works, [\pref{EVZ,EFVZ,Evans}], the multiplicative anomaly 
in the \zf\ definition
of the functional determinant has been discussed from a physical point
of view. In these calculations the anomaly arises when the field operator is 
matrix valued.
For example, for two real free scalar fields of different masses,
computing the functional determinant in two ways apparently yields different
answers. The elements of the calculation are outlined in Evans [\pref{Evans}] 
so, for
convenience, let us refer to equations (1) and (2) of this work. The classical
action is written in two ways
$$
S_a={1\over2}\int\big(\phi_1A_1\phi_1
+\phi_2A_2\phi_2\big) dx,
$$ where
$A_i=-\nabla^2+m_i^2$, and in matrix form
$$
S_b={1\over2}\int\widetilde\bPhi A\bPhi dx
$$
where 
$$
\bPhi=\left({\phi_1\atop\phi_2}\right),\quad A=\left(\matrix{
-\nabla^2+m_1^2&0\cr0&-\nabla^2+m_2^2}\right).
$$

Although trivially $S_a=S_b$, when the functional integral for the partition
function is formally
evaluated, two different answers appear. The reason given is that for
$S_a$ one naturally gets (we leave off standard factors and exponents)
$$\Det A_1\times\Det A_2
\eql{det1}$$ 
while $S_b$, gives
$$\Det A=\Det\big(A_1A_2\big)
\eql{det2}$$ 
and these are not the same. 

This last statement is certainly correct, and is a statement of the 
multiplicative anomaly. In this short note we wish to investigate, not
this mathematical anomaly, but the step leading to (\peq{det2}). This
equation appears in [\pref{EVZ,EFVZ}] where it is attributed to Benson 
{\it et al} [\pref{BBD}]  who
state it without comment. Our opinion is that this relation is not
obvious. It says that, when evaluating the functional determinant of $A$, the
finite algebraic determinant of $A$ can be taken first. The
reasons why we find this to be unnatural, and even wrong, are as follows.

Firstly, the most natural, and the most usual, way of implementing the 
\zf\ method in the
vector/matrix valued case is to take the vector index $i$ together with the
space-time coordinate $x$ as a generalised continuous index. (This has been
a standard procedure, employed most extensively by De Witt.) It leads, in
particular, to the split form (\peq{det1}).

Now, the functional integral formula for the determinant 
is an extension to the continuous, functional case of a standard finite
dimensional formula. We can check (\peq{det2}) by considering
a finite dimensional restriction. Thus replace the action by
$$
S_f={1\over2}\sum_{ij\al\be} \phi_{i\al} A_{ij\al\be}\phi_{j\be}
$$
where integers $\al$ and $\be$ play the roles of the arguments, $x$ and $y$,
of the (nonlocal) operator $A$ and have finite ranges. 
The multiple integral over the 
variables $\phi_{i\al}$ will then involve an ordinary determinant of the
matrix $A$ where the matrix indices are the pairs $(i,\al)$ and $(j,\be)$,  
and this is the correct answer.
For example, if the range of $\al$ and $\be$ is 1 to 2, then the determinant 
is a four by four one. 

The argument leading to 
(\peq{det2}) now would give
$$
\det_{\al\be}\big(\det_{ij} A_{ij\al\be}\big)
$$
\ie one takes the determinant on the $ij$ indices first and then that
on the $\al\be$ indices of the resulting expression. It is easily seen that 
these two routes give different answers. In our example, the second gives a 
sum of eight terms, each a product of four $A$ coefficients, while the 
four by four determinant expands to 24 such terms. 

Our conclusion is that if one uses the natural, and in our view correct, 
implementation of the \zf\ approach, it should not be necessary, at least in 
the vector valued case, to invoke the multiplicative anomaly, nor its 
specific expression.

\section{\bf References}
\vskip 5truept
\begin{putreferences}
\ref{APS}{Atiyah,M.F., V.K.Patodi and I.M.Singer: Spectral asymmetry and 
Riemannian geometry \mpcps{77}{75}{43}.}
\ref{AandT}{Awada,M.A. and D.J.Toms: Induced gravitational and gauge-field 
actions from quantised matter fields in non-abelian Kaluza-Klein thory 
\np{245}{84}{161}.}
\ref{BandI}{Baacke,J. and Y.Igarishi: Casimir energy of confined massive 
quarks \prD{27}{83}{460}.}
\ref{Barnesa}{Barnes,E.W.: On the Theory of the multiple Gamma function 
{\it Trans. Camb. Phil. Soc.} {\bf 19} (1903) 374.}
\ref{Barnesb}{Barnes,E.W.: On the asymptotic expansion of integral 
functions of multiple linear sequence, {\it Trans. Camb. Phil. 
Soc.} {\bf 19} (1903) 426.}
\ref{Barv}{Barvinsky,A.O. Yu.A.Kamenshchik and I.P.Karmazin: One-loop 
quantum cosmology \aop {219}{92}{201}.}
\ref{BandM}{Beers,B.L. and Millman, R.S. :The spectra of the 
Laplace-Beltrami
operator on compact, semisimple Lie groups. \ajm{99}{1975}{801-807}.}
\ref{BandH}{Bender,C.M. and P.Hays: Zero point energy of fields in a 
confined volume \prD{14}{76}{2622}.}
\ref{BBG}{Bla\v zi\' c,N., Bokan,N. and Gilkey,P.B.: Spectral geometry of the 
form valued Laplacian for manifolds with boundary \ijpam{23}{92}{103-120}}
\ref{BEK}{Bordag,M., E.Elizalde and K.Kirsten: { Heat kernel 
coefficients of the Laplace operator on the D-dimensional ball}, 
\jmp{37}{96}{895}.}
\ref{BGKE}{Bordag,M., B.Geyer, K.Kirsten and E.Elizalde,: { Zeta function
determinant of the Laplace operator on the D-dimensional ball}, 
\cmp{179}{96}{215}.}
\ref{BKD}{Bordag,M., K.Kirsten,K. and Dowker,J.S.: Heat kernels and
functional determinants on the generalized cone \cmp{182}{96}{371}.}
\ref{Branson}{Branson,T.P.: Conformally covariant equations on differential
forms \cpde{7}{82}{393-431}.}
\ref{BandG2}{Branson,T.P. and Gilkey,P.B. {\it Comm. Partial Diff. Eqns.}
{\bf 15} (1990) 245.}
\ref{BGV}{Branson,T.P., P.B.Gilkey and D.V.Vassilevich {\it The Asymptotics
of the Laplacian on a manifold with boundary} II, hep-th/9504029.}
\ref{BCZ1}{Bytsenko,A.A, Cognola,G. and Zerbini, S. : Quantum fields in
hyperbolic space-times with finite spatial volume, hep-th/9605209.}
\ref{BCZ2}{Bytsenko,A.A, Cognola,G. and Zerbini, S. : Determinant of 
Laplacian on a non-compact 3-dimensional hyperbolic manifold with finite
volume, hep-th /9608089.}
\ref{CandH2}{Camporesi,R. and Higuchi, A.: Plancherel measure for $p$-forms
in real hyperbolic space, \jgp{15}{94}{57-94}.} 
\ref{CandH}{Camporesi,R. and A.Higuchi {\it On the eigenfunctions of the 
Dirac operator on spheres and real hyperbolic spaces}, gr-qc/9505009.}
\ref{ChandD}{Chang, Peter and J.S.Dowker :Vacuum energy on orbifold factors
of spheres, \np{395}{93}{407}.}
\ref{cheeg1}{Cheeger, J.: Spectral Geometry of Singular Riemannian Spaces.
\jdg {18}{83}{575}.}
\ref{cheeg2}{Cheeger,J.: Hodge theory of complex cones {\it Ast\'erisque} 
{\bf 101-102}(1983) 118-134}
\ref{Chou}{Chou,A.W.: The Dirac operator on spaces with conical 
singularities and positive scalar curvature, \tams{289}{85}{1-40}.}
\ref{CandT}{Copeland,E. and Toms,D.J.: Quantized antisymmetric 
tensor fields and self-consistent dimensional reduction 
in higher-dimensional spacetimes, \break\np{255}{85}{201}}
\ref{DandH}{D'Eath,P.D. and J.J.Halliwell: Fermions in quantum cosmology 
\prD{35}{87}{1100}.}
\ref{cheeg3}{Cheeger,J.:Analytic torsion and the heat equation. \aom{109}
{79}{259-322}.}
\ref{DandE}{D'Eath,P.D. and G.V.M.Esposito: Local boundary conditions for 
Dirac operator and one-loop quantum cosmology \prD{43}{91}{3234}.}
\ref{DandE2}{D'Eath,P.D. and G.V.M.Esposito: Spectral boundary conditions 
in one-loop quantum cosmology \prD{44}{91}{1713}.}
\ref{BBD}{Benson,K., Bernstein,J. and Dodelson,S. \prD{44}{91}{2480}.}
\ref{Dow1}{Dowker,J.S.: Effective action on spherical domains, \cmp{162}{94}
{633}.}
\ref{Dow8}{Dowker,J.S. {\it Robin conditions on the Euclidean ball} 
MUTP/95/7; hep-th\break/9506042. {\it Class. Quant.Grav.} to be published.}
\ref{Dow9}{Dowker,J.S. {\it Oddball determinants} MUTP/95/12; 
hep-th/9507096.}
\ref{Dow10}{Dowker,J.S. {\it Spin on the 4-ball}, 
hep-th/9508082, {\it Phys. Lett. B}, to be published.}
\ref{DandA2}{Dowker,J.S. and J.S.Apps, {\it Functional determinants on 
certain domains}. To appear in the Proceedings of the 6th Moscow Quantum 
Gravity Seminar held in Moscow, June 1995; hep-th/9506204.}
\ref{DABK}{Dowker,J.S., Apps,J.S., Bordag,M. and Kirsten,K.: Spectral 
invariants for the Dirac equation with various boundary conditions 
{\it Class. Quant.Grav.} to be published, hep-th/9511060.}
\ref{EandR}{E.Elizalde and A.Romeo : An integral involving the
generalized zeta function, {\it International J. of Math. and 
Phys.} {\bf13} (1994) 453.}
\ref{ELV2}{Elizalde, E., Lygren, M. and Vassilevich, D.V. : Zeta function 
for the laplace operator acting on forms in a ball with gauge boundary 
conditions. hep-th/9605026}
\ref{ELV1}{Elizalde, E., Lygren, M. and Vassilevich, D.V. : Antisymmetric
tensor fields on spheres: functional determinants and non-local
counterterms, \jmp{}{96}{} to appear. hep-th/ 9602113}
\ref{EVZ}{Elizalde,E., Vanzo,L. and Zerbini,S. Zeta function regularisation
the multiplicative anomaly and the Wodzicki Residue. hep-th/9701060. 
{\it Comm.Math.Phys.}, to be published}
\ref{EFVZ}{Elizalde,E., Filippi,A, Vanzo,L. and Zerbini,S.  
One-loop effective potential for a charged self-interacting bosonic model 
at finite temperature with its related multiplicative anomaly. hep-th/9710171.}
\ref{Evans}{Evans,T.S. Regularization schemes and the multiplicative anomaly.
\break hep-th/9803184}

\ref{Kam2}{Esposito,G., A.Y.Kamenshchik, I.V.Mishakov and G.Pollifrone: 
Gravitons in one-loop quantum cosmology \prD{50}{94}{6329}; 
\prD{52}{95}{3457}.}
\ref{Erdelyi}{A.Erdelyi,W.Magnus,F.Oberhettinger and F.G.Tricomi {\it
Higher Transcendental Functions} Vol.I McGraw-Hill, New York, 1953.}
\ref{Esposito}{Esposito,G.: { Quantum Gravity, Quantum Cosmology and 
Lorentzian Geometries}, Lecture Notes in Physics, Monographs, Vol. m12, 
Springer-Verlag, Berlin 1994.}
\ref{Esposito2}{Esposito,G. {\it Nonlocal properties in Euclidean Quantum
Gravity}. To appear in Proceedings of 3rd. Workshop on Quantum Field Theory
under the Influence of External Conditions, Leipzig, September 1995; 
gr-qc/9508056.}
\ref{EKMP}{Esposito G, Kamenshchik Yu A, Mishakov I V and Pollifrone G.:
One-loop Amplitudes in Euclidean quantum gravity.
\prd {52}{96}{3457}.}
\ref{ETP}{Esposito,G., H.A.Morales-T\'ecotl and L.O.Pimentel {\it Essential
self-adjointness in one-loop quantum cosmology}, gr-qc/9510020.}
\ref{FORW}{Forgacs,P., L.O'Raifeartaigh and A.Wipf: Scattering theory, U(1) 
anomaly and index theorems for compact and non-compact manifolds 
\np{293}{87}{559}.}
\ref{GandM}{Gallot S. and Meyer,D. : Op\'erateur de coubure et Laplacian
des formes diff\'eren-\break tielles d'une vari\'et\'e riemannienne 
\jmpa{54}{1975}
{289}.}
\ref{Gilkey1}{Gilkey, P.B, Invariance theory, the heat equation and the
Atiyah-Singer index theorem, 2nd. Edn., CRC Press, Boca Raton 1995.}
\ref{Gilkey2}{Gilkey,P.B.:On the index of geometric operators for 
Riemannian manifolds with boundary \aim{102}{93}{129}.}
\ref{Gilkey3}{Gilkey,P.B.: The boundary integrand in the formula for the 
signature and Euler characteristic of a manifold with boundary 
\aim{15}{75}{334}.}
\ref{Grubb}{Grubb,G. {\it Comm. Partial Diff. Eqns.} {\bf 17} (1992) 
2031.}
\ref{GandS1}{Grubb,G. and R.T.Seeley \cras{317}{1993}{1124}; \invm{121}{95}
{481}.}
\ref{GandS}{G\"unther,P. and Schimming,R.:Curvature and spectrum of compact
Riemannian manifolds, \jdg{12}{77}{599-618}.}
\ref{IandT}{Ikeda,A. and Taniguchi,Y.:Spectra and eigenforms of the 
Laplacian
on $S^n$ and $P^n(C)$. \ojm{15}{1978}{515-546}.}
\ref{IandK}{Iwasaki,I. and Katase,K. :On the spectra of Laplace operator
on $\La^*(S^n)$ \pja{55}{79}{141}.}
\ref{JandK}{Jaroszewicz,T. and P.S.Kurzepa: Polyakov spin factors and 
Laplacians on homogeneous spaces \aop{213}{92}{135}.}
\ref{Kam}{Kamenshchik,Yu.A. and I.V.Mishakov: Fermions in one-loop quantum 
cosmology \prD{47}{93}{1380}.}
\ref{KandM}{Kamenshchik,Yu.A. and I.V.Mishakov: Zeta function technique for
quantum cosmology {\it Int. J. Mod. Phys.} {\bf A7} (1992) 3265.}
\ref{KandC}{Kirsten,K. and Cognola.G,: { Heat-kernel coefficients and 
functional determinants for higher spin fields on the ball} \cqg{13}{96}
{633-644}.}
\ref{Levitin}{Levitin,M.: { Dirichlet and Neumann invariants for Euclidean
balls}, {\it Diff. Geom. and its Appl.}, to be published.}
\ref{Luck}{Luckock,H.C.: Mixed boundary conditions in quantum field theory 
\jmp{32}{91}{1755}.}
\ref{MandL}{Luckock,H.C. and Moss,I.G,: The quantum geometry of random 
surfaces and spinning strings \cqg{6}{89}{1993}.}
\ref{Ma}{Ma,Z.Q.: Axial anomaly and index theorem for a two-dimensional 
disc 
with boundary \jpa{19}{86}{L317}.}
\ref{Mcav}{McAvity,D.M.: Heat-kernel asymptotics for mixed boundary 
conditions \cqg{9}{92}{1983}.}
\ref{MandV}{Marachevsky,V.N. and D.V.Vassilevich {\it Diffeomorphism
invariant eigenvalue \break problem for metric perturbations in a bounded 
region}, SPbU-IP-95, \break gr-qc/9509051.}
\ref{Milton}{Milton,K.A.: Zero point energy of confined fermions 
\prD{22}{80}{1444}.}
\ref{MandS}{Mishchenko,A.V. and Yu.A.Sitenko: Spectral boundary conditions 
and index theorem for two-dimensional manifolds with boundary 
\aop{218}{92}{199}.}
\ref{Moss}{Moss,I.G.: Boundary terms in the heat-kernel expansion 
\cqg{6}{89}{759}.}
\ref{MandP}{Moss,I.G. and S.J.Poletti: Conformal anomaly on an Einstein space 
with boundary \pl{B333}{94}{326}.}
\ref{MandP2}{Moss,I.G. and S.J.Poletti \np{341}{90}{155}.}
\ref{NandOC}{Nash, C. and O'Connor,D.J.: Determinants of Laplacians, the 
Ray-Singer torsion on lens spaces and the Riemann zeta function 
\jmp{36}{95}{1462}.}
\ref{NandS}{Niemi,A.J. and G.W.Semenoff: Index theorem on open infinite 
manifolds \np {269}{86}{131}.}
\ref{NandT}{Ninomiya,M. and C.I.Tan: Axial anomaly and index thorem for 
manifolds with boundary \np{245}{85}{199}.}
\ref{norlund2}{N\"orlund~N. E.:M\'emoire sur les polynomes de Bernoulli.
\am {4}{21} {1922}.}
\ref{Poletti}{Poletti,S.J. \pl{B249}{90}{355}.}
\ref{RandT}{Russell,I.H. and Toms D.J.: Vacuum energy for massive forms 
in $R^m\times S^N$, \cqg{4}{86}{1357}.}
\ref{RandS}{R\"omer,H. and P.B.Schroer \pl{21}{77}{182}.}
\ref{Trautman}{Trautman,A.: Spinors and Dirac operators on hypersurfaces 
\jmp{33}{92}{4011}.}
\ref{Vass}{Vassilevich,D.V.{Vector fields on a disk with mixed 
boundary conditions} gr-qc /9404052.}
\ref{Voros}{Voros,A.:
Spectral functions, special functions and the Selberg zeta function.
\cmp{110}{87}439.}
\ref{Ray}{Ray,D.B.: Reidemeister torsion and the Laplacian on lens
spaces \aim{4}{70}{109}.}
\ref{McandO}{McAvity,D.M. and Osborn,H. Asymptotic expansion of the heat kernel
for generalised boundary conditions \cqg{8}{91}{1445}.}
\ref{AandE}{Avramidi,I. and Esposito,G. Heat kernel asymptotics with 
generalised boundary conditions, hep-th/9701018.}
\ref{MandS}{Moss,I.G. and Silva P.J., Invariant boundary conditions for
gauge theories gr-qc/9610023.}
\ref{barv}{Barvinsky,A.O.\pl{195B}{87}{344}.}
\ref{krantz}{Krantz,S.G. Partial Differential Equations and Complex
Analysis (CRC Press, Boca Raton, 1992).}
\ref{treves}{Treves,F. Introduction to Pseudodifferential and Fourier Integral
Operators,\break Vol.1, (Plenum Press,New York,1980).}
\ref{EandS}{Egorov,Yu.V. and Shubin,M.A. Partial Differential Equations
(Springer-Verlag, Berlin,1991).}
\ref{AandS}{Abramowitz,M. and Stegun,I.A. Handbook of Mathematical Functions 
(Dover, New York, 1972).}
\ref{ACNY}{Abouelsaood,A., Callan,C.G., Nappi,C.R. and Yost,S.A.\np{280}{87}
{599}.}
\ref{BGKE}{Bordag,M., B.Geyer, K.Kirsten and E.Elizalde, { Zeta function
determinant of the Laplace operator on the D-dimensional ball}, 
\cmp{179}{96}{215}.}

\end{putreferences}
\bye